\documentclass[a4paper,10pt]{IEEEtran}

\usepackage{balance}

\usepackage[utf8]{inputenc}

\usepackage{balance}
\usepackage{graphicx}
\usepackage[tight,TABTOPCAP]{subfigure}
\usepackage{amsmath}
\usepackage{amssymb}
\usepackage[noend]{algorithmic}
\usepackage{algorithm}
\usepackage{listings}
\lstdefinestyle{C}{
    language=C,
    basicstyle=\ttfamily\small,
    numberblanklines=true,
    columns=fixed,
    aboveskip=2pt,
    belowskip=1pt,
    lineskip=0pt,
    numbers=left,
    numberstyle=\tiny,
    numberfirstline=true,
    firstnumber=1,
    xleftmargin=15pt,
    morekeywords={assert},
}

\newcommand{\figurerule}{\hrule width \hsize height .33pt} %%% rule definition copied from sigplanconf.cls

\usepackage{color}
\usepackage{xspace}

\usepackage[pdfpagemode=none,bookmarks=false]{hyperref}

\usepackage{stackrel}

\usepackage{multirow}

\makeatletter
\renewcommand\fs@ruled{\def\@fs@cfont{\bfseries}\let\@fs@capt\floatc@ruled
  \def\@fs@pre{\hrule height.8pt depth0pt \kern2pt}%
  \def\@fs@post{\kern2pt\hrule\relax\vspace{-2mm}}%
  \def\@fs@mid{\kern2pt\hrule\kern2pt}%
  \let\@fs@iftopcapt\iftrue}
\makeatother

\usepackage{dcolumn}
\newcolumntype{d}{D{.}{.}{5.3}}
\newcolumntype{e}{D{.}{.}{4.3}}
\newcolumntype{f}{D{.}{.}{3.3}}
\newcolumntype{g}{D{.}{.}{3.1}}
\newcolumntype{h}{D{.}{.}{2.2}}
\newcolumntype{i}{D{.}{.}{4.1}}

\newcommand{\less}{\sqsubseteq}

\newcommand{\smallsec}[1]{\smallskip\noindent{\bf #1.}}
\newcommand{\smallersec}[1]{\smallskip\noindent{\em #1.}}

\newcommand{\blast}     {{\small\sc Blast}\xspace}
\newcommand{\cpachecker}{{\small\sc CPAchecker}\xspace}
\newcommand{\cbmc}      {{\small\sc CBMC}\xspace}

\newcommand{\escjava}   {{\small\sc ESC/Java}\xspace}
\newcommand{\slam}      {{\small\sc SLAM}\xspace}
\newcommand{\jpf}       {{\small\sc JPF}\xspace}
\newcommand{\sycmc}     {{\small\sc SyCMC}\xspace}

\newcommand{\true}{\mathit{true}}
\newcommand{\false}{\mathit{false}}
\newcommand{\seq}[1]{{\langle #1 \rangle}}
\newcommand{\sem}[1]{[\![ #1 ]\!]}

\newcommand{\locs}{\mathit{L}}

\newcommand{\pc}{\mathit{pc}}

\newcommand{\cpa}{\mathbb{D}}

\newcommand{\loccpa}{\mathbb{L}}

\newcommand{\unrollcpa}{\mathbb{R}}
\newcommand{\predcpa}{\mathbb{P}}

\newcommand{\compositecpa}{\mathbb{C}}
\newcommand{\assumptionscpa}{\mathbb{A}}
\newcommand{\overflowcpa}{\mathbb{O}}

\newcommand{\Nats}{\mathbb{N}}
\newcommand{\Bools}{\mathbb{B}}
\newcommand{\Ints}{\mathbb{Z}}

\newcommand{\strengthen}{\mathord{\downarrow}}
\newcommand{\transconc}[1]{\smash{\stackrel{#1}{\rightarrow}}}
\newcommand{\transabs}[2]{\smash{\stackrel[#2]{#1}{\rightsquigarrow}}}
\newcommand{\merge}{\mathsf{merge}}
\newcommand{\stopop}{\mathsf{stop}}
\newcommand{\wait}{\mathsf{waitlist}}
\newcommand{\reached}{\mathsf{reached}}

\renewcommand{\implies}{\Rightarrow}

\newcommand{\precisions}{\Pi}

\begin{document}

\pagestyle{empty}
\begin{minipage}{17cm}
\begin{center}
~\\[1cm]
\Huge{Conditional Model Checking}
\\[2cm]
\large{Dirk Beyer\,$^{1,2}$, Thomas A. Henzinger\,$^3$, M. Erkan Keremoglu\,$^2$, Philipp Wendler\,$^1$}
\\[1cm]
\normalsize
{$^1$\,University of Passau, Germany}\\
{$^2$\,Simon Fraser University, B.C., Canada}\\
{$^3$\,IST Austria, Austria}\\[5cm]

\hspace{-5mm}
\includegraphics[scale=0.2]{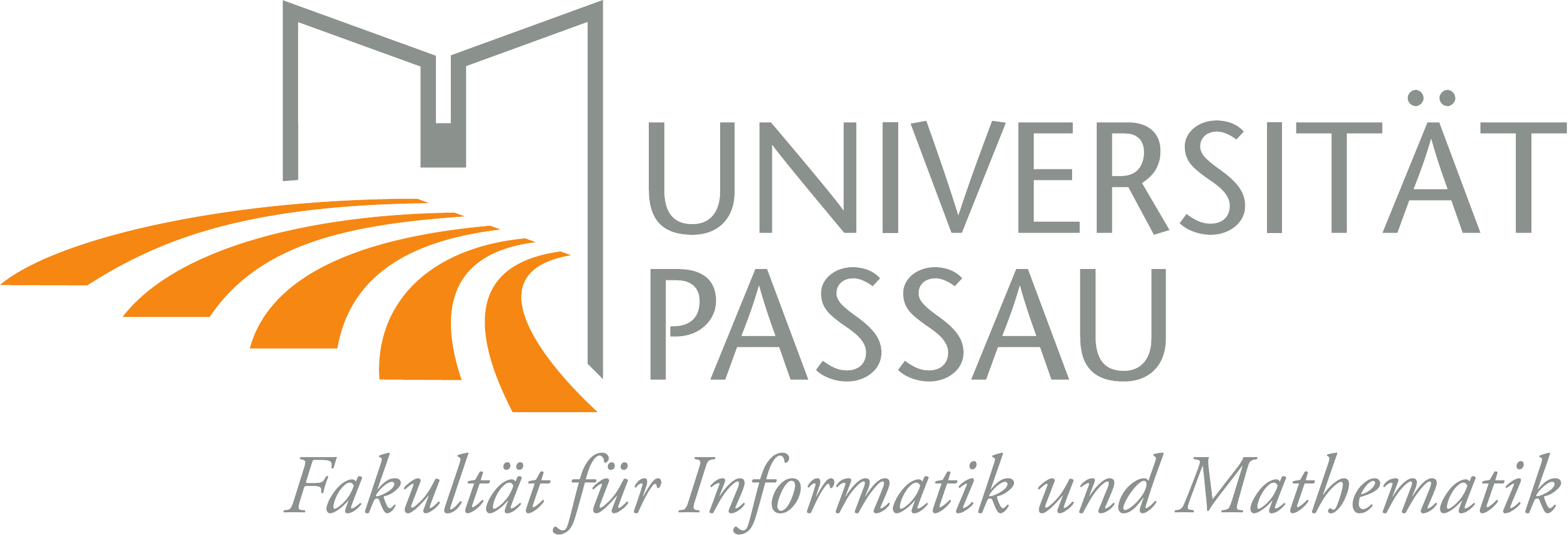} \\[1cm]
Technical Report, Number MIP-1107\\
Department of Computer Science and Mathematics\\
University of Passau, Germany\\
September 2011
\end{center}
\end{minipage}

\title{Conditional Model Checking}

\author{{Dirk Beyer\,$^{1,2}$, Thomas A. Henzinger\,$^3$, M. Erkan Keremoglu\,$^2$, Philipp Wendler\,$^1$}
\vspace{3mm}\\
{$^1$\,University of Passau, Germany}\\
{$^2$\,Simon Fraser University, B.C., Canada}\\
{$^3$\,IST Austria, Austria}\\
}

\maketitle
\thispagestyle{empty}
\setcounter{page}{1}
\pagestyle{plain}

\begin{abstract}
Software model checking, as an undecidable problem, has three possible
outcomes:
(1)~the program satisfies the specification,
(2)~the program does not satisfy the specification, and
(3)~the model checker fails.
The third outcome usually manifests itself in a space-out, time-out,
or one component of the verification tool giving up;
in all of these failing cases, significant computation is performed by the
verification tool before the failure, but no result is reported.
We propose to reformulate the model-checking problem as follows, in
order to have the verification tool report a summary of the performed
work even in case of failure:
given a program and a specification, the model checker returns a
condition~$\Psi$ ---usually a state predicate--- such that the program
satisfies the specification under the condition~$\Psi$ ---that is, as
long as the program does not leave states in which $\Psi$ is satisfied.
We are of course interested in model checkers that return conditions~$\Psi$
that are as weak as possible.
Instead of outcome~(1), the model checker will return $\Psi=\true$;
instead of~(2),
the condition~$\Psi$ will return the part of the state space that satisfies the specification;
and in case~(3), the condition~$\Psi$ can summarize the work that has
been performed by the model checker before space-out, time-out, or
giving up.
If complete verification is necessary, then a different verification
method or tool may be used to focus on the states that violate the
condition.
We give such conditions as input to a conditional model checker,
such that the verification problem
is restricted to the part of the state space that satisfies the condition.
Our experiments show that repeated application of conditional model checkers,
using different conditions,
can significantly improve the
verification results, state-space coverage, and performance.

\end{abstract}

\section{Introduction}

Model checking is an automatic search-based procedure that
exhaustively verifies whether a given model (e.g., labeled transition system)
satisfies a given specification (e.g., temporal-logic formula)~\cite{ClarkeEmerson81,QueilleSifakis82}.
For models with infinitely many states,
a model checker usually first constructs an abstract model with finitely many states.
A successful technique for this abstraction process
is counterexample-guided abstraction refinement (CEGAR)~\cite{ClarkeCEGAR},
which is an iterative process that starts with a coarse abstract model
and successively refines the abstract model,
according to information learned from abstract counterexamples.

CEGAR-based model checking was successfully applied to software programs,
where it is referred to as software model checking.
In addition to classic abstract domains, 
the most widely used abstract analysis is predicate analysis~\cite{GrafSaidi97},
which uses a set of predicates as precision of the analysis,
that is, an abstract state is a boolean combination of predicates
that occur in the given precision set.
Predicates can be obtained from a counterexample program path (CEX)
by (1) mining the predicates from the CEX via static analysis~\cite{SLAM},
(2) interpolation along path formulas representing the CEX~\cite{AbstractionsFromProofs},
or (3) loop-invariant generation for path programs~\cite{PathPrograms}.
Further improvements of efficiency and effectiveness of the predicate analysis
can be obtained by integrating all components of the CEGAR-procedure
into one single on-the-fly algorithm for the construction of an
abstract reachability tree~\cite{LazyAbstraction}, % \cite{BLAST} 
by dynamically adjusting the analysis precision~\cite{CPAplus},
and by encoding large program blocks into 
formulas that represent many transitions~\cite{LBE}.

Since software model checking is an undecidable problem, 
there are three possible outcomes of the analysis process:
(1) the program satisfies the specification,
(2) the program does not satisfy the specification, and
(3) the model checker fails.
The first outcome can be obtained by the model checker
if the abstract model that was computed for the program
is sufficient to prove the program correct under the given specification.
This outcome can be accompanied by a proof certificate~\cite{HJMN+02}.
The second outcome can be obtained by the model checker if
an abstract counterexample path is found and can be proven
feasible, i.e., a bug that can actually occur in the program.
This outcome is usually accompanied by the violating program part
in the form of program source code, and sometimes test input
to reproduce the error at run-time~\cite{BLAST-test}.
The third outcome usually occurs if the model checker runs
out of resources (memory exhausted, time-out)
or if one of the components in the verification tool gives up.
The failure of refinement, i.e., the model checker does not succeed
in eliminating an infeasible CEX, is an example for the latter case
(this occurs if the chosen technique for predicate extraction is not strong enough).
In all of these failing cases, significant computation is performed by the
verification tool before the failure.
But since no useful result is reported, the spent resources are wasted.

With our approach of \emph{conditional model checking}, 
we propose a new definition of the model-checking outcome.
The goal of conditional model checking is to maximize the
outcome of a model-checking run under certain conditions, e.g., a given set of resources.
We reformulate the model-checking problem as follows:
Given a program and a specification, conditional model checking returns a
condition~$\Psi$ ---usually a state predicate--- such that the program
satisfies the specification under the condition~$\Psi$ ---that is, as
long as the program does not leave states in which $\Psi$ is satisfied.

\begin{figure}
\includegraphics[width=\linewidth]{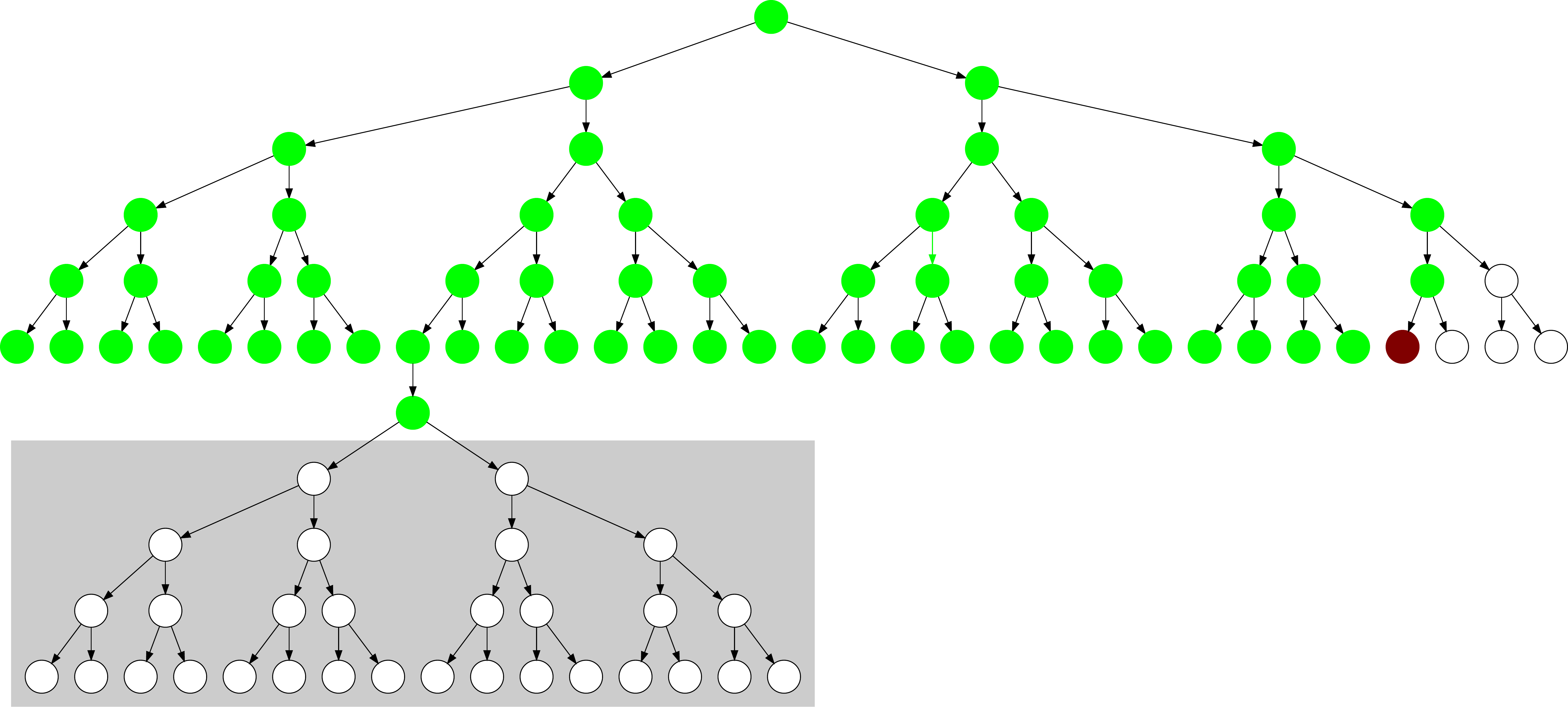}
\caption{Example state space; with condition \textit{Path Length} and limit~7,
states in the gray box are skipped by the analysis, the dark-red error state (on the right) can be found fast,
green states are verified}
\label{fig:example-art}
\vspace{-3mm}
\end{figure}

We are interested in model checkers that return conditions
$\Psi$ that are as weak as possible.
This way, the verification tool reports a summary of the performed
work even if the tool cannot completely verify the program.
The outcomes of a model-checking run can be translated to conditions in the following way: 
Previous outcome~(1) corresponds to the condition $\Psi=\true$.
That is, if the model checker returns~$\true$ as condition,
the model checker completely verified the program under no additional conditions. 
For outcome~(2), the model checker does not return~$\false$,
but can specify program parts that are free of errors,
and explicitly exclude the violating parts.
For example, consider an invalid program that consists of two branches,
one of which is safe and the other violates the specification.
The resulting condition for this program would contain
all program states that occur in the successfully verified branch
and report an error path that violates the specification.
In outcome~(3), in which the model checker previously failed with ``no useful results'',
the condition~$\Psi$ now summarizes the work that has
been performed by the model checker before space-out, time-out, or
giving up.
For example, consider a valid program that consists of two branches,
one of which is easy to verify by the model checker
and the other leads the model checker into an infinite loop.
Using conventional model checking, we would not get any feedback from the 
model checker, because the time-out would cause a failure.
Our modified approach of conditional model checking would still 
be unable to prove that the program satisfies the specification.
However, it would heuristically detect the hopeless situation
and summarize the performed work by reporting
that (at least) the first branch has been successfully verified.
If complete verification is necessary, then a different verification
method or tool may be used to focus on the states that violate the condition.
Figure~\ref{fig:example-art} illustrates an example state-space,
where the verification engine is restricted to verify only 
paths that are not longer than seven nodes.

We implemented several techniques to prevent the model checker from failure.
First, we monitor the progress for every critical component of the model checker.
When such a component fails to deliver results (exhausts its assigned time or memory), 
the monitor discovers the problem, terminates the component, and generates a condition that
excludes the corresponding states from the verification result.
Second, we implemented several conditions that try to predict situations
in which the model checker should not continue to try verifying this part of the program,
and generate conditions excluding that part.
Such conditions make it possible to verify larger parts of the program,
instead of spending all time on a particular, unsolvable problem.
Thus, if we cannot completely verify a program,
we at least obtain a ``verification coverage'' that is as large as possible,
and we can use the resulting condition to continue the verification 
with different tools and configurations.
Third, we can use the negation of the condition generated by one analysis
as the condition for a subsequent analysis,
letting the latter check those parts of the state space that the former did not verify successfully.

Our experiments show that these conditions can also significantly improve the
performance;
for example, programs for which the model checker previously failed after spending minutes,
can now be proved to violate the specification within seconds.

\smallsec{Contributions and Applications}
Conditional model checking enables the following features that were not possible
in conventional model checking before:
\begin{itemize}
  \item \emph{No Fail.} 
        Every run of the model checker results in a condition formula
        that summarizes the achieved results. 
  \item \emph{Maximal Outcome.}
        Conditional model checking maximizes the outcome for a given
        set of resources (memory/time limit).
  \item \emph{Regression Checking.} 
        The conditions can later be used for re-verification,
        i.e., if started with the condition formula as additional input,
        the conditional model checker can a~priori exclude certain parts from the 
        verification.
        If the program was only slightly changed, then this approach
        speeds up the re-checking significantly.
  \item \emph{Partial Verification.} 
        Conditional model checking can be used to restrict the verification
        to certain parts of the program, by taking as input a condition formula
        that excludes the parts that should not be verified.
        For example, some parts of the system might be checked via model checking,
        others via testing, theorem proving, or complementary model checkers.
        % Also, the ``depth of exploration'' (and thus the effort spent) could 
        % be specified as input condition.
  \item \emph{Improved Performance for Bug Hunting.}
        Conditional model checking enables the specification of conditional iteration orders.
        For example, we can specify that the model checker searches each 
        abstract path for a given amount of time (conditional DFS) or 
        each abstract path up to a certain length (similar to bounded model checking).
        This approach significantly improves the performance for programs with errors. 
  \item \emph{Benchmark Generation.} 
        Conditional model checking can be used to produce hard verification benchmarks,
        by generating excluding conditions for all parts that can (currently) not
        be model checked.
        If started with the generated conditions as input,
        the model checker will verify the maximal verifiable program fragment,
        and the time to prove this fragment correct can be measured.
  \item \emph{Comparison of Tools.} 
        Given two model checking tools, we can not only compare the
        time and memory needed for a given verification task (cf. benchmarks above),
        but we can now compare the quality of the verification results
        (the weaker the condition, the better).
\end{itemize}

\smallsec{Tool Implementation}
We implemented conditional model checking using standard components of
the open-source verification framework \cpachecker~\cite{CPACHECKER}.
Our extension of \cpachecker and all benchmark programs that we used in our experimental evaluation
are available from the supplementary website.

\begin{figure}
\figurerule
\begin{lstlisting}[style=C,firstline=3]
void main() {
  int x = 1;
  if (nondet_int()) {
    while (x < 10000) {
      x++;
    }
    assert (x == 10000);
  } else {
    x = 0;
  }
  assert (x != 0);
}
\end{lstlisting}
\caption{Example program with loop}
\label{fig:example-loop}
\vspace{-4mm}
\end{figure}

\smallsec{Example}
Loops introduce challenges for static program analysis.
For example, if the domain of abstract predicates with lazy abstraction is used for the analysis, there is the possibility that
each step of loop unwinding will add a new predicate and verification will be performed until the entire path is
unwound.
In some cases that operation might be repeated thousands of times and this will lead the analysis to get stuck in the
loop.
The example in Fig.~\ref{fig:example-loop} presents a case where the analysis might fail to terminate.
If the analysis has to fully unwind the loop in this program, the 
analysis will not terminate in a reasonable amount of time.
In that case, the verification outcome would be `fail'.
At line~11, there is another assertion and the analysis would miss the opportunity to investigate
this part of the specification because it is busy with checking the loop.
If the analysis is started with a condition that limits the number of unwindings 
of a loop to at most $k$ iterations,
it will finally skip the loop and verify the rest of the program.
This way it can determine that the assertion at line~11 does not hold and report an error without much effort.
If the specification were not violated in the rest of the program,
the outcome would be `safe' under the assumption
that the program visits the loop entry at most $k$~times
(as in bounded model checking).

\begin{figure*}
\begin{minipage}[t]{0.32\linewidth}
\centering
\includegraphics[scale=0.3]{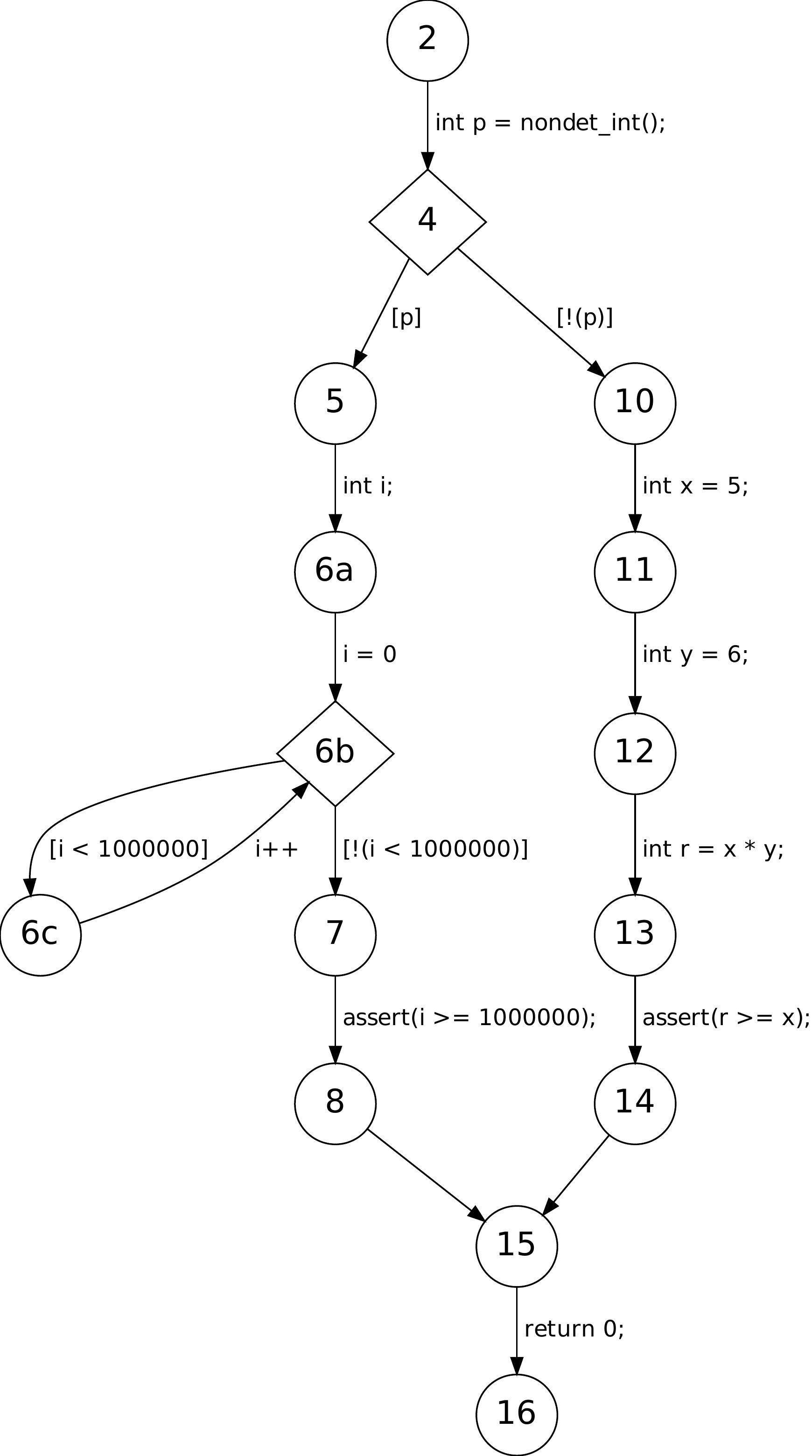}
\caption{Control-flow automaton for example program with non-linear safety condition}
\label{fig:example-non-linear-CFA}
\end{minipage}
\hfill
\begin{minipage}[t]{0.32\linewidth}
\centering
\includegraphics[scale=0.3]{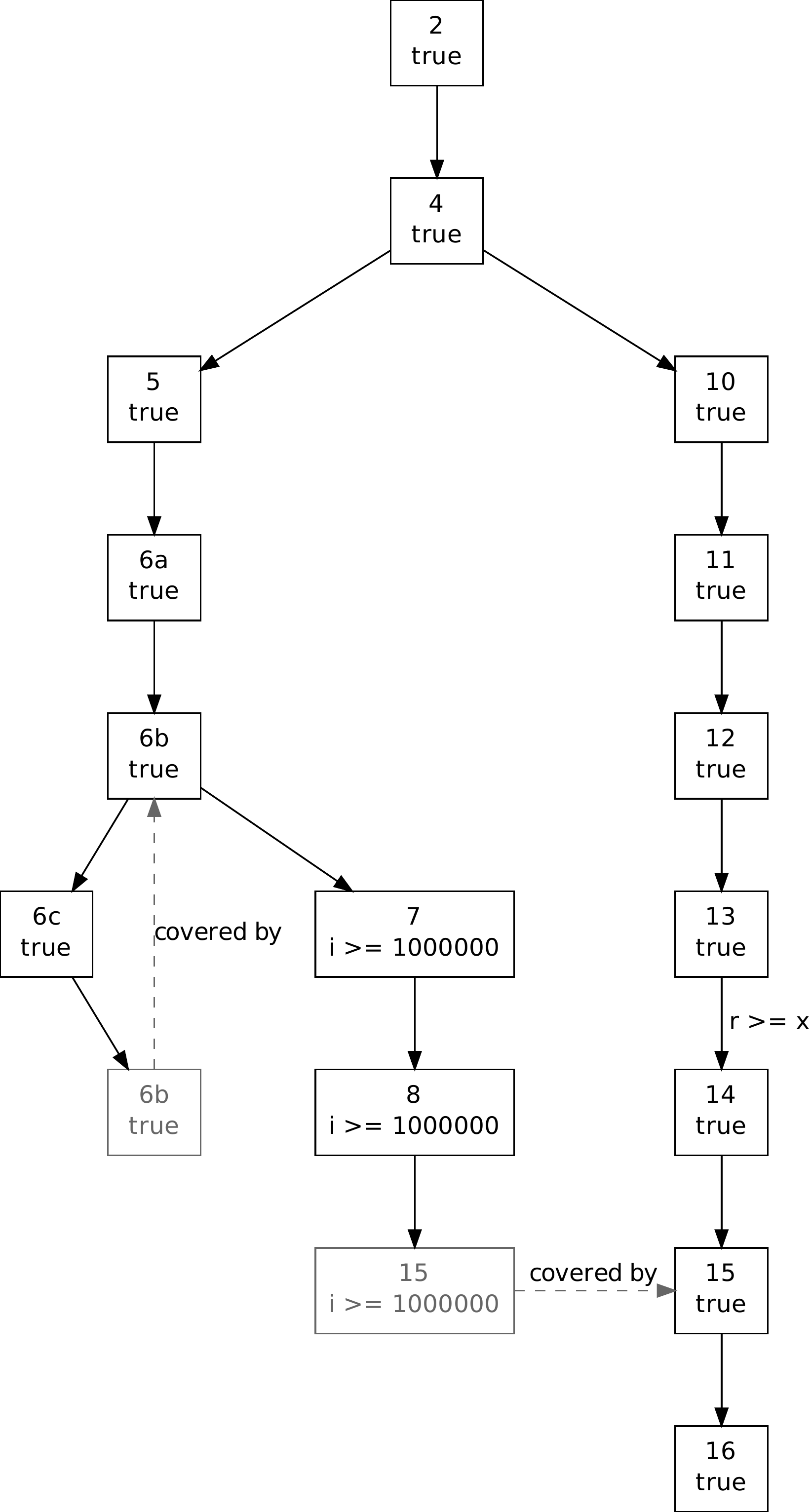}
\caption{ART for example program with non-linear safety condition; covered states are gray, edges are labeled with assumptions}
\label{fig:example-non-linear-ART}
\end{minipage}
\hfill
\begin{minipage}[t]{0.32\linewidth}
\centering
\includegraphics[scale=0.3]{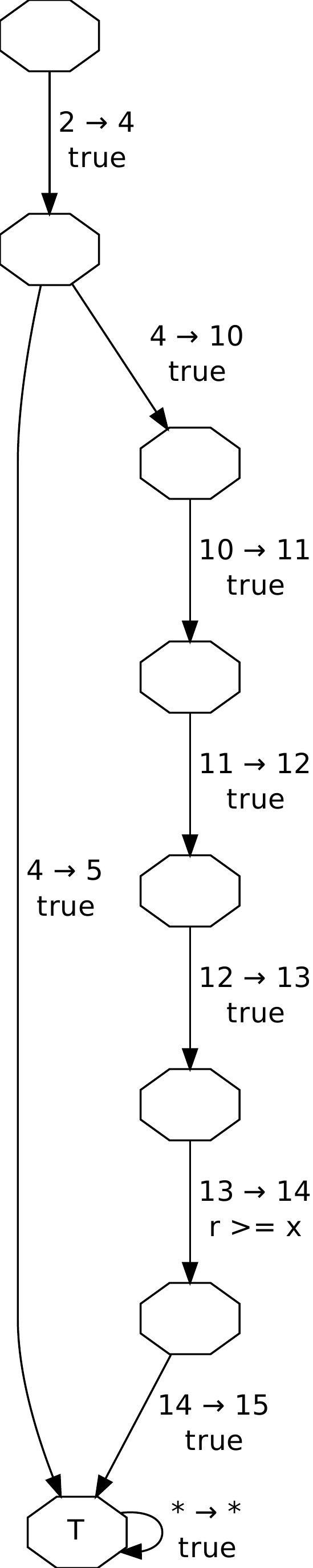}
\caption{Assumption automaton for example program with non-linear safety condition}
\label{fig:example-non-linear-Automaton}
\end{minipage}
\vspace{-4mm}
\end{figure*}

Model checking with predicate analysis depends on the capabilities of SMT solvers,
and thus can only verify conditions that can be expressed in the theories
that are supported by the integrated SMT solver.
This excludes, for example, properties that depend on non-linear relations
between program variables, as shown in the snippet of Fig.~\ref{fig:example-non-linear}.
Suppose we analyze this program with a conventional predicate analysis.
Given a precision that includes the predicate $(i \geq 1000000)$,
the model checker will easily be able to prove that the assertion in
line~7 can never fail.
However, as our predicate analysis is based on linear arithmetics
and needs to model the multiplication of program variables as uninterpreted function,
the model checker cannot prove that the second assertion in line~13 also always holds.
A precise path analysis reveals that the path is infeasible, thus,
the analysis has to give up.
Fig.~\ref{fig:example-non-linear-ART} shows the abstract reachability tree~(ART)
that the analysis would generate.
Each ART~node is labeled with the control-flow node that it belongs to and the formula
that represents the abstract state.
The ART~edges are labeled with the assumptions that the analysis
produced during the verification run.
In this case, there is only one non-trivial assumption at the edge from
program location~13 to~14.
For all other edges the assumption is $true$ and not shown in the ART.

\begin{figure}
\figurerule
\begin{lstlisting}[style=C,firstline=3]
int main() {
  int p = nondet_int();

  if (p) {
    int i;
    for (i = 0; i < 1000000; i++);
    assert(i >= 1000000);

  } else {
    int x = 5;
    int y = 6;
    int r = x * y;
    assert(r >= x);
  }
  return 0;
}
\end{lstlisting}
\caption{Example program with non-linear safety condition}
\label{fig:example-non-linear}
\vspace{-4mm}
\end{figure}

In conditional model checking, the model checker
outputs an assumption that explicitly states
which parts of the program are safe and which parts are not checked.
This assumption can be written in several formats;
we choose the following two.
First, we create an assumption formula over the program locations and variables.
In this case, the assumption would be $(\pc = 13) \implies (r \geq x)$.
This is both human readable and useful as input for a subsequent analysis.
Second, we convert the abstract reachability tree
that was produced by the model checker into an assumption automaton that is annotated
with the generated assumptions.
The assumption automaton that the analysis outputs for this example
is shown in Fig.~\ref{fig:example-non-linear-Automaton}.
Each transition of the automaton is labeled with a control-flow edge
(e.g., ``2~$\rightarrow$~4'' is the edge from location~2 to~4)
and the assumption that was used when taking this edge.
Parts of the ART where the assumption is always $true$
are collapsed into a single sink state~`T',
which has a self edge that matches any control-flow edge.
For all other parts of the ART there is a 1:1 correspondence
between ART nodes and automaton states.

Compared with the assumption formula,
the assumption automaton offers an easier way to specify complex facts over program paths
instead of just over program locations.
It is equally useful as input condition for a second analysis of the program,
for example, using a configuration that is better suitable to verify
the remaining parts of the state space.
This second analysis has to analyze only those paths of the program
that contain at least one edge where the assumption is not $true$.
This can be implemented easily by letting the automaton run in parallel to the analysis,
taking the matching transition whenever the analysis takes a program edge.
Then the exploration of a path can be stopped as soon as the automaton reaches the sink state~`T',
as this means that the assumption of all future edges would always be $true$.

For the program in Fig.~\ref{fig:example-non-linear},
a good choice for verifying the assertion in line~13
is an explicit-value analysis for integers.
If the assumption automaton from Fig.~\ref{fig:example-non-linear-Automaton}
is used to restrict the analyzed state space,
the explicit-value analysis needs to prove only the safety of the second assertion,
which it can check efficiently.
The final analysis correctly results in reporting that the program is safe.
However, note that an explicit-value analysis cannot verify the other assertion
(it would have to unroll the loop a million times,
probably exceeding all available time or memory resources).
Thus it is indeed necessary to use the assumption
from the first run to guide the second analysis.

By using two different model-checking configurations,
and by giving detailed information about the verified state space in form of a condition,
this example can be proved safe.
No configuration is able to verify this alone;
both would either fail due to resource exhaustion or 
terminate without useful results.

\smallsec{Related Work}
The assume-guarantee paradigm is a well-known principle 
of verification theory~\cite{AssumeGuarantee}.
Conditional model checking (CMC) implements this paradigm:
``if the program fulfills the (generated) assumptions, 
then the program is guaranteed to satisfy the specification''.
A formal system should explicitly state under which conditions 
it guarantees correctness.
\escjava uses an annotation language for Java to let the user encode
conditions for pruning false alarms~\cite{ESCJava}.
Thus, the user can choose a compromise between soundness and 
efficiency.
CMC follows this principle by allowing the model checker to
take conditions as input.
Bounding the path length in symbolic execution is a well-known technique~\cite{King76}.
Bounded model checking (BMC) is successful in finding bugs,
but is often rejected as verification technique
because it is unsound~\cite{BMC}.
If the condition of unwinding every loop up to a certain bound
is stated explicitly, then it would be a sound conditional model-checking technique.
The bounded model checker \cbmc does report whether the program
contains paths that exceed the given bounds,
or whether the program could be verified completely~\cite{CBMC}.
This can be nicely expressed as a condition,
but conditional model checking is more powerful as more information is reported.
Nimmer and Ernst efficiently extract program specifications from 
dynamic analysis~\cite{Nimmer}.
Although the approach is unsound, the generated specifications 
can effectively cover a large part of the state space.
Conditional model checking also has the objective to
improve the verification coverage.
Conway et al.\ define a points-to analysis with conditional soundness~\cite{Conway}.
Many software-verification tools make implicit assumptions about the program,
sacrificing soundness in order to be practically useful.
For example, the predicate-abstraction--based tools \slam and \blast 
implicitly use the assumption that the program does not contain integer overflows,
and model variables as unbounded mathematical integers (cf.~\cite{BLAST}, page~515).
Several verification tools use heuristics similar to our conditions
(search strategies, iteration orders, pruning, etc.)
in order to find safety violations faster~\cite{Dwyer2007,Musuvathi2004}.
\jpf is a tool which allows arbitrary user-defined search strategies~\cite{JPFHeuristics}.
Conditional model checking makes these heuristics externally visible and adjustable in the form of conditions.

\smallsec{Summary}
Conventional model checkers may fail without any useful result; in fact,
this is by far the most common output of model checkers in practice.
In conditional model checking, the model checker is required to 
always provide a result to the user ---
ideally a proof or counterexample, but if this is not possible,
then the conditional model checker reports the work it has done
in the form of a summary, namely, as a condition that characterizes the part of the state
space that has been proven correct.
This report can be used to improve future verification runs, by the same or by different tools, 
avoiding to repeat work that has already been done.

\section{Preliminaries}

We are briefly providing some basic notions and concepts
from the literature~\cite{BLAST,CPA},
which our definitions are based on.

\smallsec{Programs}
We restrict the presentation to
a simple imperative programming language,
where all operations are either assignments or assume operations,
and all variables range over integers.%
\footnote{Our implementation is based on \cpachecker{}~\cite{CPACHECKER},
which accepts C~programs given in C Intermediate Language {\sc Cil}~\cite{CIL},
and supports interprocedural program analysis.}
A \emph{program} is represented by a \emph{control-flow automaton} (CFA),
which consists of 
a set~$\locs$ of program locations (models the program counter~$\pc$),
an initial program location~$\pc_0$ (models the program entry), and 
a set $G \subseteq \locs \times Ops \times \locs$ of control-flow edges 
(models the operation that is executed when
control flows from one program location to another).
The set of program variables that occur in operations from~$Ops$
is denoted by~$X$.
A \emph{concrete state} of a program is 
a variable assignment $c: X \cup \{\pc\} \to \Ints$ 
that assigns to each variable an integer value.
The set of all concrete states of a program is denoted by~$C$.
A set~$r \subseteq C$ of concrete states is called a \emph{region}.
Each edge~$g \in G$ defines a (labeled) transition relation 
$\mathord{\transconc{g}} \subseteq C \times \{g\} \times C$.
The complete transition relation~$\transconc{}$ is the union over 
all control-flow edges:
$\mathord{\transconc{}} = \bigcup_{g \in G} \transconc{g}$.
We write $c \transconc{g} c'$ if $(c, g, c') \in \mathord{\transconc{}}$,
and $c \transconc{} c'$ if there exists a $g$ with $c \transconc{g} c'$.
A concrete state~$c_n$ is \emph{reachable} from a region~$r$, denoted by $c_n \in Reach(r)$, if
there exists a sequence of concrete states $\seq{c_0, c_1, \ldots, c_n}$
such that $c_0 \in r$ and for all $1 \leq i \leq n$,
we have~$c_{i-1} \transconc{} c_{i}$.

\smallsec{Configurable Program Analysis}
We formalize our reachability analysis using the framework of 
\emph{configurable program analysis} (CPA)~\cite{CPA}.
A CPA specifies ---independently of the analysis algorithm---
the abstract domain and a set of operations that control the program analysis.
Such a CPA can be plugged in as a component into the software-verification framework
without the need to work on program parsers, exploration algorithms, and
their general data structures.
A CPA
$\mathbb{C} = (D, \leadsto, \merge, \stopop)$
consists of
an abstract domain~$D$, 
  a transfer relation~$\leadsto$ (which computes abstract successor states),
  a merge operator~$\merge$ (which specifies if and how to merge abstract states when control flow meets),
  and
  a stop operator~$\stopop$ (which specifies if an abstract state is covered by another abstract state).
The abstract domain~$D = (C, {\cal E}, \sem{\cdot})$ consists of 
  a set~$C$ of concrete states,
  a semi-lattice~${\cal E}$ over abstract-domain elements, and
  a concretization function that maps each abstract-domain element to the represented 
    set of concrete states.

Using this framework, program analyses can be composed of several component CPAs.
For example, a standard predicate analysis can be created as the composition
of the following two CPAs.

\smallsec{CPA for Location Analysis}
The CPA for \emph{location analysis} 
$\loccpa = (D_\loccpa, \transabs{}{}_\loccpa,
             \merge_\loccpa, \stopop_\loccpa)$
tracks the program counter~$\pc$ explicitly.

\smallsec{1}
  The domain $D_\loccpa$ is based on the flat lattice for the
  set~$\locs$ of program locations:\\
  $D_\loccpa = (C, \mathcal{E}_\loccpa, \sem{\cdot})$, with
    $\mathcal{E}_\loccpa = ((\locs \cup \{\top\}), \less)$,
  $l \less l'$ if $l=l'$ or $l'=\top$,
  $\sem{\top} = C$,
  and for all~$l$ in~$\locs$,
  $\sem{l} = \{ c \in C \mid c(\pc) = l \}$.

\smallsec{2}
  The transfer relation $\transabs{}{}_\loccpa$ has the transfer
  $l \transabs{g}{}_\loccpa l'$ ~if~
  $g = (l,\cdot,l')$.

\smallsec{3}
  The merge operator does not combine elements when control flow meets:
  $\merge_\loccpa(l, l') = l'$

\smallsec{4}
  The termination check returns true
  if the current element is already in the reached set:
  $\stopop_\loccpa(l, R) = (l \in R)$.

\smallsec{CPA for Predicate Analysis}
\label{sec:predabscpa}
The CPA for \emph{predicate analysis} 
$\predcpa = (D_\predcpa, \transabs{}{}_\predcpa, 
             \merge_\predcpa, \stopop_\predcpa)$,
a program analysis for predicate abstraction
that tracks the validity of a finite set~$\precisions$ of predicates over program variables,
consists of the following components:

\smallsec{1}
The domain $D_\predcpa = (C, \mathcal{E}, \sem{\cdot})$ 
models abstract states as formulas that represent regions (of concrete states).
The semi-lattice~$\mathcal{E} = (\mathcal{F}, \Rightarrow)$
is based on the set~$\mathcal{F}$ of quantifier-free boolean formulas
over variables from~$X$.
The concretization function~$\sem{\cdot}: \mathcal{F} \to 2^C$
assigns to each abstract state~$e$ its meaning, i.e., 
the set of concrete states that it represents:
$\sem{e} = \{ c \in C \mid c \models e \}$.

\smallsec{2}
The transfer relation $\transabs{}{}_\predcpa$ has the transfer
$e \transabs{g}{}_\predcpa e'$ if
$e'$ is the strongest boolean combination of predicates from~$\precisions$
that is implied by the strongest postcondition of the abstract state~$e$ and the operation of~$g$.

\smallsec{3}
The merge operator does not combine elements when control flow meets:
$\merge_\predcpa(e, e') = e'$.

\smallsec{4}
The termination check considers abstract states individually:
$\stopop_\predcpa(e, R) = (\exists e' \in R: e \implies e')$.

\begin{algorithm}[b]
\begin{small}
\caption{\textbf{ }$\textit{CPA}(\cpa, R_0, W_0)$
         \label{algo:analysis}}
\begin{algorithmic}[1]
\INPUT               a CPA $\cpa = (D, \transabs{}{}, \merge, \stopop)$,\\
\hspace{3.7mm}       a set~$R_0 \subseteq E$ of abstract states,\\ 
\hspace{3.7mm}       a subset~$W_0 \subseteq R_0$ of frontier abstract states,\\
\hspace{3.7mm}          where $E$ denotes the set of elements of the semi-lattice of $D$
\OUTPUT              a set of reachable abstract states,\\
\hspace{5.6mm}      a subset of frontier abstract states
\VARDECL             two sets $\reached$ and $\wait$ of elements of $E$

\STATE $\reached := R_0$;
\STATE $\wait := W_0$;
\WHILE {$\wait \not= \emptyset$}
  \STATE choose $e$ from $\wait$; remove $e$ from $\wait$; 
  \FOR{each $e'$ with $e \transabs{}{} e'$}
    \FOR{each $e'' \in \reached$}
      \STATE // Combine with existing abstract state.
      \STATE $e_{new} := \merge(e', e'')$;
      \IF {$e_{new} \not= e''$}
        \STATE $\wait    := \big(\wait    \cup \{e_{new}\}\big) \setminus \{e''\}$;
        \STATE $\reached := \big(\reached \cup \{e_{new}\}\big) \setminus \{e''\}$;
      \ENDIF
    \ENDFOR
    \STATE // Add new abstract state?
    \IF {$\lnot~\mathord{\stopop}(e', \reached)$}
      \STATE $\wait := \wait \cup \{e'\}$;
      \STATE $\reached := \reached \cup \{e'\}$;
    \ENDIF
  \ENDFOR
\ENDWHILE
\STATE {\bf return } $(\reached, \emptyset)$
\end{algorithmic}
\end{small}
\end{algorithm}

\smallsec{Analysis Algorithm}
Algorithm~\ref{algo:analysis} shows the program-analysis algorithm that
is implemented in the tool \cpachecker.
The algorithm gets as input a CPA and two sets of abstract states:
one is the set~$R_0$ ($\reached$) of reached abstract states, and 
one is the set~$W_0$ ($\wait$) of abstract states that the algorithm is told to process next.
The algorithm stops if the set $\wait$ is empty (all abstract states completely processed)
and returns the two sets~$\reached$ and~$\wait$.

For the first call of the algorithm, the parameter arguments will be singletons
containing the initial abstract state.
In each iteration of the `\lstinline{while}' loop, 
the algorithm takes one state~$e$ from the waitlist,
computes all abstract successors and processes each
of them as~$e'$.

Next, the algorithm checks (lines 6--11) if there is an existing abstract state in the 
set of reached states with which the new state is to be merged
(e.g., at join points where control flow meets after completed branching).
If this is the case, then the new, merged abstract state is substituted 
for the existing abstract state in both sets $\reached$ and $\wait$.
(This operation is sound because the merge operation is not allowed to under-approximate.)

In lines 12--15, the stop operator ensures that the new abstract state is inserted into the work sets
only if it is a new state, i.e., not covered by a state that is already in the set~$\reached$.

\section{Conditional Model Checking}

In the following, we characterize common failure causes for model checkers,
present preventing conditions for those causes,
and later give formal definitions for the program analyses that we
implemented to experiment with conditional model checking.

\subsection{Failure Classes and Failure-Preventing Conditions}
\label{sec:conditions}

\smallsec{Classes of Failures}
There are various reasons for a model checker to fail during the verification process.
We are particularly interested in classifying the failures according to the 
part of the verification algorithm where the problem occurs.

\begin{enumerate}
  \item \textbf{Global Progress.} 
        If the model checker does not stop continuously adding new abstract states to the set
        of reachable states, it will sooner or later run out of resources.
        For example, in an explicit-value analysis, a loop might be unfolded many times 
        (cycling) and many explicit values are stored for a particular location.
        To prevent the model checker from running into such a situation,
        we can measure the number of abstract states in the reached set for a certain location 
        and generate an excluding assumption.
        Similarly, we can monitor the total number of abstract states, total memory consumption 
        and total time consumption.
  \item \textbf{Post Computation.}
        The computation of the abstract successor might fail because the
        new abstract state might be too large, or too difficult to compute (memory, time).
        If this occurs, the monitoring component stops the analysis of the specific part,
        adds an excluding assumption, and lets the analysis continue on another path. 
        Various conditions can be used to prevent the model checker from spending
        time on difficult program parts.
        For example, we could measure the size of the analysis precision
        (e.g., the number of predicates for a certain location exceeds a given threshold,
         which makes in turn the computation of the predicate abstraction expensive)
        and stop exploring the current successor if a certain threshold is exceeded.
  \item \textbf{Counterexample Analysis.}
        If an abstract error path has been found, it needs to be
        analyzed for feasibility and possible refinements.
        This step might fail in a predicate-abstraction-based analysis
        if the feasibility check of the path (performed by an external SMT solver) 
        fails, or if the refinement process did not succeed in eliminating 
        the infeasible error path (e.g., the used predicate-extraction technique 
        might not be strong enough).
        If a failure of a component occurred, an excluding assumption is generated 
        for a pivot node on the path.
        To prevent the refinement component from failure, we can measure
        the length of the path, the predicates involved, the structure of the path formulas, etc.
\end{enumerate}

\begin{figure}
\newcommand\timef{\mathit{time}}
\newcommand\mem{\mathit{mem}}
\newcommand\refin{\mathit{ref}}
\newcommand\loc{\mathit{loc}}
\newcommand\state{\mathit{state}}
\centering
\begin{tabular}{@{}l@{\hspace{1.0em}}l@{\hspace{0.5em}}l@{\hspace{0.5em}}r@{}}
Component        & Name                        & Condition                & \makebox[0pt][r]{Impl.}\\
\hline
\multirow{2}{0.2\linewidth}{Global Progress}
                 & Total Time                  & $\timef$                 & \checkmark\\
                 & Total Space                 & $\mem$                   & \checkmark\\
                 & \# Abstract States          & $|\reached|$             & \checkmark\\
                 & \# Abstract States per Loc. & $\#(\reached, \loc)$     & \\
                 & Busy Edge                   & $\#(\mathit{edge})$      & \checkmark\\
\hline
\multirow{2}{0.2\linewidth}{Post Computation}
                 & Time for Post               & $\timef(\transabs{}{})$  & \checkmark\\
                 & Space for Post              & $\mem(\transabs{}{})$ \\
                 & Size of State               & $\mem(\state)$ \\
                 & Path Length                 & $length(path)$           & \checkmark\\
                 & Time Spent in Path          & $\timef(path)$           & \checkmark\\
                 & Repeating Locs. in Path     & $\#(path, \loc)$         & \checkmark\\
                 & Assume Edges in Path        & $\mathit{assumes(path)}$ & \checkmark\\
\hline
\multirow{2}{0.2\linewidth}{CEX Analysis}
                 & Time for Refinement         & $\timef(\refin)$         & \checkmark\\
                 & Space for Refinement        & $\mem(\refin)$ \\
                 & Size of Path Formula        & $\mem(\mathit{pf})$      & \checkmark\\
\end{tabular}
\caption{Example Conditions for Conditional Model Checking;
  column \textit{Impl.} lists the conditions we implemented}
\label{fig:conditions}
\vspace{-4mm}
\end{figure}

\smallsec{Preventing Conditions}
We classify the conditions into the above-mentioned categories
according to the location of occurrence in the analysis algorithm,
and provide an incomplete list of example conditions
(as an overview, cf.~Fig.~\ref{fig:conditions}):

\smallersec{Global Progress --- Total Time and Space}
Measure the total time and memory consumption of the model checker.
If a resource limit is reached, then
the analysis is stopped and excluding assumptions are generated for all abstract states
that are still in the waitlist at this moment.
In contrast to `hard' resource limits that we can set (e.g., with {\tt ulimit}),
these `soft' limit enables the model checker to perform post-processing and
output the verification results as an assumption formula.

\smallersec{Global Progress --- Number of Abstract States}
Measures the total number of the elements in the set of reached abstract states.
If the number of states is larger than a certain threshold,
the analysis is stopped and excluding assumptions are generated for all abstract states
that are still in the waitlist at this moment.

\smallersec{Global Progress --- Number of Abstract States per Location}
Measures the number of elements in the set of reached abstract states
that belong to each program location.
If any location has more states attached to it than the limit,
all post computations for incoming edges of this location are not performed anymore,
and instead excluding assumptions are added.

\smallersec{Global Progress --- Busy Edge}
Measures the total number of post operations that the currently considered edge
of the control-flow automaton was involved in.
If the edge is used more often than allowed by a given threshold,
post operations for this edge are not performed anymore,
but rather excluding assumption are computed.

\smallersec{Post Computation --- Time and Space for Post}
Measures the time of each transfer in a component analysis.
If the operation takes longer than the specified time limit or exceeds the specified
memory limit, the operation is terminated, and an excluding assumption for the
abstract state is added.

\smallersec{Post Computation --- Size of Abstract State}
Measures the size of the current abstract state, and 
if the specified threshold is reached, an excluding assumption is added.
If predicate analysis is used, we take the number of disjuncts
in the abstraction formula of the abstract predicate state.

\smallersec{Post Computation --- Path Length}
Measures the number of abstract states on the path to the current abstract state, and 
if the specified threshold is reached, an excluding assumption is added.
(If several paths lead to the current abstract state, the maximum value is considered.)

\smallersec{Post Computation --- Time Spent in Path}
Measures the total amount of time consumed on the path to the current abstract state, and
if the specified time limit is exceed, an excluding assumption is added.
(If several paths lead to the current abstract state, the maximum value is considered.)

\smallersec{Post Computation --- Repeating Locations in Path} 
Measures the number of occurrences of the current program location on the path
from the root to the current abstract state in the abstract reachability tree.
If a program location is encountered more than a specified number of times on a single path,
an excluding assumption is added.

\smallersec{Post Computation --- Assume Edges in Path}
Measures the total number of assume edges seen on the path to the current abstract state, and
if the specified threshold is exceed, an excluding assumption is added.
(If several paths lead to the current abstract state, the maximum value is considered.)

\smallersec{Counterexample Analysis --- Time and Space for Refinement}
The execution of each refinement step for a given analysis is monitored.
If the operation takes longer than the specified time limit or exceeds the specified
memory limit, the operation is terminated, and an excluding assumption for the
abstract state is added.

\smallersec{Counterexample Analysis --- Size of Path Formula}
We measure the size of the path formula, which is used for checking feasibility
and for computing interpolants for predicate refinement.
If the formula exceeds a certain size, we do not start
the SMT solver on that formula, but rather add an excluding assumption.

\subsection{Formalization and Implementation}

\begin{figure}
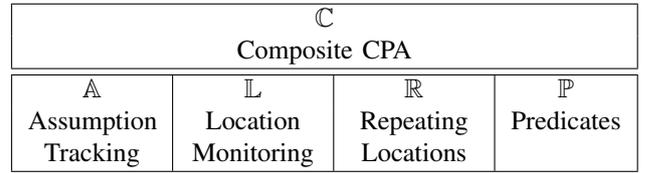

\centering
\newcolumntype{P}{p{0.19\linewidth}}
\newcolumntype{C}{>{\centering}P}
\begin{tabular}{|C|C|C|P|}
\hline
\multicolumn{4}{|c|}{$\compositecpa$} \\
\multicolumn{4}{|c|}{Composite CPA} \\\hline\hline
$\assumptionscpa$   & $\loccpa$           & $\unrollcpa$         & \multicolumn{1}{c|}{$\predcpa$ }\\
Assumption Tracking & Location Monitoring & Repeating Locations  & \multicolumn{1}{c|}{Predicates} \\\hline
\end{tabular}
\caption{Schematic view of an exemplary CPA combination}
\label{fig:cpa-structure}
\vspace{-4mm}
\end{figure}

We formalize \emph{conditional model checking}
using the framework of configurable program analysis (CPA)~\cite{CPAplus}.
We re-use existing CPAs for domains like predicate 
and location analysis,
and add new CPAs for tracking conditions and assembling assumptions.
Then, we use a composite CPA to combine them.
An example for a CPA configuration structure is illustrated in Figure~\ref{fig:cpa-structure}.

\smallsec{CPA for Assumptions}
The CPA for \emph{Assumption Storage} $\assumptionscpa = (D_\assumptionscpa, \transabs{}{}_\assumptionscpa, \merge_\assumptionscpa, \stopop_\assumptionscpa)$
is a CPA that stores the assumptions collected during the analysis.
The abstract domain~$D_\assumptionscpa$,
the transfer relation~$\transabs{}{}_\assumptionscpa$,
the merge operator~$\merge_\assumptionscpa$,
and the stop operator~$\stopop_\assumptionscpa$
are explained in the following.

\smallsec{1}
The domain~$D_\assumptionscpa = (C, \mathcal{E}_\assumptionscpa, \sem{\cdot}_\assumptionscpa)$ is based on
the semi-lattice $\mathcal{E}_\assumptionscpa = (\mathcal{F}, \Rightarrow)$, where
$\mathcal{F}$~is the set of quantifier-free formulas over variables from~$X$.
An abstract state represents an assumption made by the model checker about
the concrete states of the program.
The meaning of the assumption is that the verification result 
is guaranteed if all assumptions are fulfilled.

\smallsec{2}
The transfer relation~$\transabs{}{}_\assumptionscpa$ is the set
$\{ (e, g, e')\ |\ e \neq \false, \linebreak g\in G, e'=\true\}$.
The successor state is always the abstract state that makes no assumptions.
This will be refined by the strengthen operator~$\strengthen$ of the composite~CPA.
If the predecessor state has the assumption~$\false$,
the analysis does not proceed because the assumptions already exclude the current path.

\smallsec{3}
The merge operator computes the conjunction of both assumptions:
$\merge_\assumptionscpa(e, e') = e\wedge e'$.
Merging the abstract states of~$\assumptionscpa$ ensures that the number
of abstract states is not increased by adding assumptions to the analysis
(as there will never be two abstract states in the reached set
which differ only in their assumptions).

\smallsec{4}
The termination stop considers abstract states individually:
$\stopop_\assumptionscpa(e, R) = (\exists e' \in R: e' \implies e)$
returns true, if there is a state in the set of reached states
that has a stricter assumption.

\smallsec{CPA for Overflow Monitoring}
Variables of a computer program are always bounded (e.g., to 32 bits)
and therefore have a minimum as well as a maximum value.
If the value exceeds the maximum, an overflow occurs,
i.e., the program continues with a different value.
However, predicate analyses usually ignore overflows 
and model variables as unbounded mathematical integers. 
Thus, the analysis would be unsound.

We define a CPA that observes the progress and generates the necessary assumptions
under which the analysis is sound.
These assumptions can also be used to remove infeasible paths
from the explored state space.
The \emph{CPA for Overflow Monitoring}
$\overflowcpa = (D_\overflowcpa, \transabs{}{}_\overflowcpa,
             \merge_\overflowcpa, \stopop_\overflowcpa)$
tracks the assumptions for
overflows of program variables.
For simplicity, we assume that all program variables have the same type
with a minimal value of $\mathsf{MIN}$ and a maximum value of $\mathsf{MAX}$.

\smallsec{1}
The domain $D_\overflowcpa$ is based on boolean formulas:
$D_\overflowcpa = (C, \mathcal{E}_\overflowcpa, \sem{\cdot})$,
with $\mathcal{E}_\overflowcpa = (\mathcal{F}, \Rightarrow)$,
where $\mathcal{F}$ is the set of quantifier-free boolean formulas over variables from~$X$.

\smallsec{2}
The transfer relation creates an assumption if the current program operation
might lead to an overflow.
If the edge $g$ is an assignment of the form $x = t$,
where $x\in X$ and $t$~is a term over variables from $X$,
the successor state $\psi'$ is the formula $(x \geq \mathsf{MIN}) \wedge (x \leq \mathsf{MAX})$.
In other cases the successor state is $\true$.
We assume that overflows occur only at the end of the computation,
when the result of~$t$ is assigned to~$x$.
Statements in which overflows occur in intermediate results
are transformed into simpler statements by pre-processing.

If this CPA produces an assumption, it will be added to the abstract state
of the CPA for predicate analysis by the strengthen operator~$\strengthen$.
This will help the analysis by excluding parts of the state space
that are infeasible because of an overflow for a variable.

\smallsec{3}
The merge operator combines elements by taking the conjunction:
$\merge_\overflowcpa(\psi, \psi') = \psi\wedge \psi'$.

\smallsec{4}
The termination stop $\stopop_\overflowcpa$ always returns $\true$.

\smallsec{CPA for Repeating Locations Monitoring}
We present a CPA that monitors how often a single location was encountered
on a path and stops the analysis when this number exceeds a threshold.
The \emph{CPA for repeating locations monitoring} 
$\unrollcpa = (D_\unrollcpa, \transabs{}{}_\unrollcpa, 
             \merge_\unrollcpa, \stopop_\unrollcpa)$
tracks the maximum number of times a location has
been seen on a path leading to the current abstract state,
and stops the analysis when appropriate.

\smallsec{1}
  The domain $D_\unrollcpa$ is based on
  a map from locations
  to natural numbers:
  $D_\unrollcpa = (C, \mathcal{E}_\unrollcpa, \sem{\cdot})$, with
    $\mathcal{E}_\unrollcpa = (((\locs \to \Nats) \times \Bools), \less)$,
  $(e, t) \less (e', t')$ if
  $t = t'$ and
  for all~$l_e$ in~$\locs$, $e(l_e) = e'(l_e)$.
  %%% \top needed?

  An abstract state consists of (1) a map from locations to natural numbers
  that stores how often a location has been seen on the path to the current
  state, and (2) a flag that shows whether at least one location was seen
  more often than the user-specified threshold $k$ for repetition of locations in a path.

\smallsec{2}
  The transfer relation $\transabs{}{}_\unrollcpa$ has the transfer
  $(e, t) \transabs{g}{}_\unrollcpa ((e', t))$ for
  $g = (l,\cdot,l_s)$ ~if~
      (1) $t' = (\exists l'' \in L : e'(l'') > k)$
  and (2) the following holds:
          \[ \forall l''\in L :  e'(l'') = \begin{cases} e(l'') + 1 & \text{ if }l'' = l_s \\ e(l'') & \text{ else }\end{cases}\]

\smallsec{3}
  The merge operator combines elements by taking the maximum
  number of occurrences for each location when control flow meets:
  $\merge_\unrollcpa((e, t), (e', t')) = (e'', t'')$ such that
  for all~$l$ in~$\locs$, we have
  $e''(l) = \max(e(l),e'(l))$ and $t'' = t \vee t''$.

\smallsec{4}
  The termination check always returns $\true$
  because the decision whether an abstract state is covered
  does not depend on the state of the conditions CPA.

\smallsec{Composite CPA}
The composite CPA combines an existing CPA like
the predicate abstraction CPA~$\predcpa$
with the assumption tracking CPA $\assumptionscpa$,
the CPA for location monitoring $\loccpa$,
and one CPA for conditions that limits the search space.
These CPAs are called {\em component CPAs}.
The definition of the composite CPA can easily be extended
in order to have several condition CPAs run in parallel
by adding more component CPAs.

As an example we present here the composite CPA for
predicate abstraction together with monitoring of repeating locations.
Other combinations are similar.
In fact, only a single operator (the strengthen operator~$\strengthen$)
and the merge strategy
need to be changed in order to adopt the CPA to other configurations.

\smallsec{1}
  The domain $D_\compositecpa$ is defined by the cross product
  of the domains of $\assumptionscpa$, $\loccpa$, $\unrollcpa$ and $\predcpa$.

\smallsec{2}
  The transfer relation $\transabs{}{}_\compositecpa$ is the cross product
  of the transfer relations of the composite CPAs,
  but with the strengthen operator $\strengthen$ applied on the successor states:
  \begin{align*}
  \transabs{}{}_\compositecpa = \{&
     ((e_\assumptionscpa, e_\loccpa, e_\unrollcpa, e_\predcpa),
      g,
      (e''_\assumptionscpa, e''_\loccpa, e''_\unrollcpa, e''_\predcpa),
      )
     \mid \\
    & (e_\assumptionscpa, g, e'_\assumptionscpa) \in \transabs{}{}_\assumptionscpa,
      (e_\loccpa, g, e'_\loccpa) \in \transabs{}{}_\loccpa, \\
    & (e_\unrollcpa, g, e'_\unrollcpa) \in \transabs{}{}_\unrollcpa,
      (e_\predcpa, g, e'_\predcpa) \in \transabs{}{}_\predcpa, \\
    & (e''_\assumptionscpa, e''_\loccpa, e''_\unrollcpa, e''_\predcpa) = \strengthen(e'_\assumptionscpa, e'_\loccpa, e'_\unrollcpa, e'_\predcpa)
    \}
  \end{align*}

  In the case the conditions CPA detects that a threshold is exceeded
  (for repeating locations monitoring this means that $e'_\unrollcpa = (\cdot, \true)$),
  the strengthen operator adds an assumption to the abstract state
  that excludes the current path:
  $\strengthen(e'_\assumptionscpa, e'_\loccpa, e'_\unrollcpa, e'_\predcpa) =
    (\false, e'_\loccpa, e'_\unrollcpa, e'_\predcpa)$ if $e'_\unrollcpa = (\cdot, \true)$.
  Otherwise it is the identity function.
  If another CPA that generates assumptions (like the Overflow Monitoring CPA) is present,
  the strengthen operator would also add the assumptions generated by this CPA to the
  assumptions stored by the Assumption Storage CPA,
  and to the abstract state of the Predicate Abstraction CPA.

\smallsec{3}
  The merge operator $\merge_\compositecpa$ merges elements,
  if the location and the predicate abstraction part of them is equal:
  \vspace{-1mm}
  \begin{multline*}
    \merge((e_\assumptionscpa, e_\loccpa, e_\unrollcpa, e_\predcpa), (e'_\assumptionscpa, e_\loccpa, e'_\unrollcpa, e_\predcpa)) \\
    = (\merge_\assumptionscpa(e_\assumptionscpa, e'_\assumptionscpa), e_\loccpa,
       \merge_\unrollcpa(e_\unrollcpa, e'_\unrollcpa), e_\predcpa)
  \end{multline*}
  Otherwise, it returns its second argument (i.e., does not merge).

  This relies on the fact that $\merge_\predcpa$ never merges abstract states.
  If another CPA with a different merge implementation is used
  instead of $\predcpa$ this needs to be adjusted.

\smallsec{4}
  The stop operator is the conjunction of the stop operators of the component CPAs:
  $\stopop((e_\assumptionscpa, e_\loccpa, e_\unrollcpa, e_\predcpa), R)
    = \stopop_\assumptionscpa(e_\assumptionscpa, R_\assumptionscpa) \wedge
       \stopop_\loccpa(e_\loccpa, R_\loccpa) \wedge
       \stopop_\unrollcpa(e_\unrollcpa, R_\unrollcpa) \wedge
       \stopop_\predcpa(e_\predcpa, R_\predcpa)$

  where $R_\assumptionscpa$, $R_\loccpa$, $R_\unrollcpa$ and $R_\predcpa$ are the projections
  of $R$ to the respective parts of the tuple that form an abstract state.

\smallsec{Assumptions Post-Processing}
Once the model checker decides to terminate 
(either verification is complete, a violation is found, or resource limit reached)
some post-processing is performed in order to produce the assumption formula
as the result of the model checker run.
The result of the $CPA$ algorithm is the tuple $(\reached, \wait)$.
For each abstract state $(e_\assumptionscpa, l, \cdot, e_\predcpa) \in \reached$
that is either still contained in the waitlist
or belongs to the error location (i.e., $l = l_e$),
the formula $(\pc = l) \Rightarrow \neg e_\predcpa$ is created.
For all other states $(e_\assumptionscpa, l, \cdot, e_\predcpa) \in \reached$
the formula $(\pc = l) \implies (\neg e_\predcpa \lor e_\assumptionscpa)$ is created.
Then, the final invariant is the conjunction of all these formulas.

This can be extended to more complex analyses like CEGAR.
If the model checker detects an infeasible error path $(e_0, \ldots, e_n)$
but fails to compute a better precision that would exclude this path,
the assumption $(\pc = l_i) \Rightarrow \neg {e_\predcpa}_i$ is added
for all states~$e_i = (\cdot, l_li, \cdot, {e_\predcpa}_i)$ beginning
with the first unreachable state in the path until $e_n$.

As an additional output, the model checker always returns the 
abstract reachability tree~\cite{BLAST},
which serves as proof certificate in case of successful verification,
and contains the error path in case a violation of the specification is found
(a human-readable version of the error path is also printed).

\section{Experimental Results}

Our experiments give evidence of the practical benefits of conditional model checking over 
conventional model checking.
We evaluate the following three aspects:

First, we generalize the statement that bounded model checking is effective and efficient 
for finding bugs to conditional model checking:
not only the number of loop unwindings can restrict the state space in such a way that shallow bugs can easily be found, but also other conditions can be effective for this.
For example, we use conditional model checking with conditions on the path length.

Second, we show how to improve the verification result by applying 
different complementary tools or analysis approaches,
one after the other, leveraging the complementary strengths of the different approaches.
Restricting the resources that are spent for a particular analysis approach can
dramatically improve the overall verification performance
(i.e., the total sum of consumed analysis resources)
and also the effectiveness (number of solved problems).

Third, we show that providing the conditions produced by one verification approach
(which did not succeed to completely verify the program) to another approach or tool,
can improve the performance and effectiveness of successive verification runs.

\smallsec{Configurations}
We experimented with several verification configurations of the tools \cbmc and \cpachecker.
The tool \cbmc is a bounded model checker that 
is based on a tightly integrated SMT solver for precise bitwise arithmetics~\cite{CBMC}.
This tool is extremely fast in finding shallow bugs,
but generally cannot prove safety of programs that contain 
loops with many iterations or with unknown bounds.
It is configured with a parameter~$k$
that limits the number of unwindings of the loops in the program.
The tool \cpachecker is an open verification platform with support for explicit-value
analysis and predicate analysis.
We implemented the conditions from Sect.~\ref{sec:conditions}
and apply some of them to condition the verification engine of \cpachecker.
We configure \cpachecker to perform an explicit-value analysis, 
which keeps track of integer values explicitly
while searching in depth-first order through the state space.
In this configuration, the abstract states at meet points of the control flow
are not joined in order to obtain a precise analysis.
However, such a precise analysis usually exhausts the given memory quickly for large programs,
and may get stuck in loops since no abstractions or summaries are computed.
We also configure \cpachecker for a predicate analysis with lazy abstraction~\cite{LazyAbstraction},
CEGAR~\cite{ClarkeCEGAR}, and large-block encoding (LBE)~\cite{LBE}.
This is a powerful but expensive analysis, consuming large amounts of resources 
when verifying large programs.
It is based on linear arithmetics in order to allow for Craig interpolation,
and therefore cannot handle overflows or bitwise operators.
In both configurations, \cpachecker was configured to check every error report using \cbmc
(i.e., whenever \cpachecker finds an error path, it generates a C program for the path and
 queries \cbmc about the feasibility of it).
In cases where CBMC determines a path as infeasible,
we continue the analysis in order to check for the existence of a different feasible error path.
However, if no error path is found, the analysis result is `unknown'
instead of `program is safe'
(because there might be some abstract states
that are covered by the infeasible error path which were not analyzed).

\smallsec{Experimental Setup and Reporting}
All experiments were performed on a machine with a 3.4\,GHz Quad Core CPU and 16\,GB of RAM.
The operating system was Ubuntu~10.10 (64~bit), using Linux~2.6.35 as kernel and
OpenJDK~1.6 as Java virtual machine.
We took the \cpachecker components from revision~3820 of the repository and \cbmc in version\,4.0 for running the experiments.
Unless stated otherwise, a hard time limit of 15~minutes and a memory limit of 14\,GB were used.
\cpachecker was configured with a Java heap size of 4\,GB when using predicate abstraction
(to leave enough RAM for the SMT solver),
and 12\,GB for other configurations.
This was sufficient memory for all experiments.

For each run, the tables show the consumed processor time 
of the verifier (in seconds) and a symbol for the result.
The symbol~`\checkmark' indicates that the model checker computed the correct answer
(bug found or safe).
A dash means that the model checker did not succeed in completely checking the program.
There was no experiment for which a verifier computed a wrong answer 
(no false-positives, no true-negatives).
Programs whose name contains {\sc bug} are known to be unsafe.

\smallsec{Benchmark Programs}
We ran our experimental setup on three sets of benchmark verification problems.
The first set contains simplified, partial Windows~NT device drivers;
the second set contains simplified versions of the state machine that
handles the communication in the SSH suite.
Different numbers in a program name indicate different
safety properties that are verified.
These two sets of benchmark programs were taken from the \blast repository\,%
\footnote{
\href{http://www.sosy-lab.org/~dbeyer/Blast/}{{\footnotesize\tt http://www.sosy-lab.org/$\sim$dbeyer/Blast}}}.
The third set of programs contains SystemC programs from the supplementary web page
of \sycmc~\cite{SYCMC}\,%
\footnote{
\href{https://es.fbk.eu/people/roveri/tests/fmcad2010/}{{\footnotesize\tt https://es.fbk.eu/people/roveri/tests/fmcad2010}}}.
\subsection{Model Checking using Conditions}

\begin{table*} 
\centering
\normalsize
\newcommand\smallerfontsize\scriptsize
\begin{tabular}{l|l|D{.}{.}{3.0}c|lD{.}{.}{2.1}c}
&				& \multicolumn{2}{c|}{Conventional MC}	& \multicolumn{3}{c}{Conditional MC}\\
Configuration	& Program	& \multicolumn{2}{c|}{Result}	& Condition \& Threshold		& \multicolumn{2}{c}{Result} \\
\hline
\multirow{5}{1.2cm}{Explicit Analysis}
&\verb|pc_sfifo_1_BUG|		& 180 & -			& Repeating Locations in Path: 3	&  1.5  & \checkmark \\
&\verb|pc_sfifo_2_BUG|		& 180 & -			& Repeating Locations in Path: 3	&  1.6  & \checkmark \\
&\verb|test_locks_14.BUG|	& 180 & -			& Path Length: \phantom{6}85		& 33    & \checkmark \\
&\verb|test_locks_15.BUG|	& 180 & -			& Path Length: \phantom{6}90		& 54    & \checkmark \\
&\verb|transmitter.15.BUG|	& 180 & -			& Path Length: 600			&  4.9  & \checkmark \\
&\verb|transmitter.16.BUG|	& 180 & -			& Path Length: 600			&  3.3  & \checkmark \\
\end{tabular}
\caption{Conventional versus conditional analysis on programs with bug}
\label{tab:experiments-buggy}
\vspace{-10mm}
\end{table*}

First we show that conditional model checking (by applying state-space restriction conditions)
can significantly increase the verification effectiveness and reduce the verification resources
when applied to the problem of finding program bugs.

\smallsec{Configurations}
Table~\ref{tab:experiments-buggy} reports performance results for explicit analysis
for programs with a known bug.
The time limit was set to 3\,min for the explicit analysis since it is an effective
technique to detect easy-to-find bugs and a small time limit is usually enough
for this configuration to terminate with a result.
The column `Condition \& Threshold' lists the used condition and 
the corresponding threshold value.
For example `Path Length: 90' means that the analysis verified
only paths up to a length of 90 program statements.

\smallsec{Discussion}
Conventional model checking fails to find the violation of the specification within the
specified time limit.
Conditional model checking using explicit analysis with the condition `Path Length',
or condition `Repeating Locations in Path'
identifies the error in the examples.
The experiments show a significant performance improvement:
The two {\small \verb|pc_sfifo|} benchmark programs, which were
analyzed using Repeating Locations in Path condition with a threshold value of 3, were falsified in less than
2\,s.
The {\small \verb|test_locks_*.BUG|} programs were tested using Path Length condition
with threshold values 85 and 90 respectively, and were falsified in less than 1\,min.
For the {\small \verb|transmitter.*.BUG|} programs the same condition
was used with a threshold of 600, and they were falsified in less than 5\,s.

\subsection{Repeated Model Checking with Different Conditions}
\label{sec:restarts}

We demonstrate that different tools and different configurations of conditional
model checkers can significantly improve the results of the verification process,
by systematically starting the model checkers with different verification objectives.

\newcommand\T{\rule{0pt}{1.9ex}}
\newcommand\B{\rule[-1.2ex]{0pt}{0pt}}

\begin{table*}
\centering
\footnotesize
\newcommand\smallerfontsize\tiny
\begin{tabular}{l |D{.}{.}{1.2}c |D{.}{.}{4.2}c |D{.}{.}{3.1}c |D{.}{.}{4.1}c |D{.}{.}{4.1}c |D{.}{.}{4.1}c |D{.}{.}{4.1}c}
& \multicolumn{4}{c|}{\cbmc} & \multicolumn{6}{c|}{\cpachecker} & \multicolumn{2}{c|}{Comb. A} & \multicolumn{2}{c}{Comb. B} \\
& & & & & \multicolumn{2}{c|}{Explicit} & \multicolumn{2}{c|}{Explicit} & \multicolumn{2}{c|}{Predicate}& \multicolumn{2}{c|}{Expl. + Pred.} & \multicolumn{2}{c}{CBMC+Expl.+Pred.} \\
& \multicolumn{2}{c|}{$k=1$} & \multicolumn{2}{c|}{$k=10$} & \multicolumn{2}{c|}{$time (10s)$} & & & & & & & & \\
\hline \T
\verb|cdaudio_simpl1|     & 1.0  & \checkmark & 1.0 & \checkmark & 3.5 & \checkmark & 3.2 & \checkmark & 17  & \checkmark  & 3.5 & \checkmark & 1.0 & \checkmark \\
\verb|cdaudio_simpl1_BUG| & .99  & \checkmark & .99 & \checkmark & 2.8 & \checkmark & 2.5 & \checkmark & 10  & \checkmark  & 2.8 & \checkmark & .99 & \checkmark \\
\verb|diskperf_simpl1|    & .25  & -          & .26 & -          & 13  & -          & 900 & -          & 15  & \checkmark  & 28  & \checkmark & 28  & \checkmark \\
\verb|floppy_simpl3|      & .16  & \checkmark & .15 & \checkmark & 2.9 & \checkmark & 2.4 & \checkmark & 8.5 & \checkmark  & 2.9 & \checkmark & .16 & \checkmark \\
\verb|floppy_simpl3_BUG|  & .16  & \checkmark & .16 & \checkmark & 2.2 & \checkmark & 2.5 & \checkmark & 7.8 & \checkmark  & 2.2 & \checkmark & .16 & \checkmark \\
\verb|floppy_simpl4|      & .28  & \checkmark & .28 & \checkmark & 3.5 & \checkmark & 3.7 & \checkmark & 12  & \checkmark  & 3.5 & \checkmark & .28 & \checkmark \\
\verb|floppy_simpl4_BUG|  & .30  & \checkmark & .30 & \checkmark & 3.0 & \checkmark & 2.9 & \checkmark & 11  & \checkmark  & 3.0 & \checkmark & .30 & \checkmark \\
\verb|kbfiltr_simpl1|     & .05  & \checkmark & .06 & \checkmark & 3.1 & \checkmark & 2.2 & \checkmark & 3.7 & \checkmark  & 3.1 & \checkmark & .05 & \checkmark \\
\verb|kbfiltr_simpl2|     & .11  & \checkmark & .10 & \checkmark & 3.1 & \checkmark & 3.2 & \checkmark & 5.2 & \checkmark  & 3.1 & \checkmark & .11 & \checkmark \\
\verb|kbfiltr_simpl2_BUG| & .11  & \checkmark & .12 & \checkmark & 2.2 & \checkmark & 2.1 & \checkmark & 4.1 & \checkmark  & 2.2 & \checkmark & .11 & \checkmark \\
\hline \T \B \bfseries NT drivers total & 3.4 & 9 & 3.4 & 9 & 39 & 9 & 920 & 9 & 94 & 10 & 54 & 10 & 31 & 10\\\hline\hline \T 
\verb|s3_clnt_1|          & .03  & -          & 4.3 & -          & 9.8 & \checkmark & 8.3 & \checkmark & 8.1 & \checkmark & 9.8 & \checkmark & 9.8 & \checkmark \\
\verb|s3_clnt_1_BUG|      & .03  & -          & 4.2 & -          & 4.4 & \checkmark & 3.6 & \checkmark & 6.2 & \checkmark & 4.4 & \checkmark & 4.4 & \checkmark \\
\verb|s3_clnt_2|          & .03  & -          & 4.6 & -          & 10  & \checkmark & 8.2 & \checkmark & 7.2 & \checkmark  & 10  & \checkmark & 10  & \checkmark \\
\verb|s3_clnt_2_BUG|      & .03  & -          & 4.3 & -          & 4.5 & \checkmark & 3.5 & \checkmark & 5.4 & \checkmark  & 4.5 & \checkmark & 4.5 & \checkmark \\
\verb|s3_clnt_3|          & .03  & -          & 5.3 & -          & 9.7 & \checkmark & 8.1 & \checkmark & 5.8 & \checkmark  & 9.7 & \checkmark & 9.7 & \checkmark \\
\verb|s3_clnt_3_BUG|      & .03  & -          & 5.3 & \checkmark & 5.1 & \checkmark & 3.5 & \checkmark & 6.0 & \checkmark  & 5.1 & \checkmark & 5.1 & \checkmark \\
\verb|s3_clnt_4|          & .03  & -          & 4.8 & -          & 9.8 & \checkmark & 8.5 & \checkmark & 10  & \checkmark  & 9.8 & \checkmark & 9.8 & \checkmark \\
\verb|s3_clnt_4_BUG|      & .03  & -          & 4.3 & -          & 4.3 & \checkmark & 3.5 & \checkmark & 6.4 & \checkmark  & 4.3 & \checkmark & 4.3 & \checkmark \\
\verb|s3_srvr_1|          & .03  & -          & 4.1 & -          & 3.3 & \checkmark & 2.4 & \checkmark & 21  & \checkmark  & 3.3 & \checkmark & 3.3 & \checkmark \\
\verb|s3_srvr_1_BUG|      & .03  & -          & 6.4 & \checkmark & 2.2 & \checkmark & 1.7 & \checkmark & 4.8 & \checkmark  & 2.2 & \checkmark & 2.2 & \checkmark \\
\verb|s3_srvr_2|          & .03  & -          & 5.5 & -          & 2.8 & \checkmark & 2.4 & \checkmark & 150 & \checkmark  & 2.8 & \checkmark & 2.8 & \checkmark \\
\verb|s3_srvr_2_BUG|      & .03  & -          & 6.2 & \checkmark & 1.8 & \checkmark & 1.7 & \checkmark & 4.1 & \checkmark  & 1.8 & \checkmark & 1.8 & \checkmark \\
\verb|s3_srvr_3|          & .03  & -          & 5.6 & -          & 3.6 & -          & 2.6 & -          & 9.0 & \checkmark & 13  & \checkmark  & 13  & \checkmark \\
\verb|s3_srvr_4|          & .03  & -          & 5.6 & -          & 3.1 & -          & 2.5 & -          & 28  & \checkmark  & 31  & \checkmark & 31  & \checkmark \\
\verb|s3_srvr_6|          & .03  & -          & 6.6 & -          & 14  & -          & 250 & \checkmark & 230 & \checkmark  & 240 & \checkmark & 240 & \checkmark \\
\verb|s3_srvr_7|          & .03  & -          & 6.3 & -          & 14  & -          & 200 & \checkmark & 47  & \checkmark & 61  & \checkmark  & 61  & \checkmark \\
\verb|s3_srvr_8|          & .03  & -          & 6.1 & -          & 2.8 & \checkmark & 3.0 & \checkmark & 23  & \checkmark & 2.8 & \checkmark  & 2.8 & \checkmark \\
\hline \T \B \bfseries SSH total & 0.51 & 0 & 90 & 3 & 110 & 13 & 510 & 15 & 570 & 17 & 420 & 17 & 420 & 17\\\hline\hline \T 
\verb|bist_cell|          & .02  & -          & .65 & -          & 9.4 & -          & 9.2 & -          & 210 & \checkmark  & 220 & \checkmark & 220 & \checkmark \\
\verb|kundu|              & .02  & -          & 120 & -          & 7.3 & \checkmark & 6.2 & \checkmark & 900 & -           & 7.3 & \checkmark & 7.3 & \checkmark \\
\verb|kundu1_BUG|         & .02  & -          & 18  & -          & 2.0 & \checkmark & 1.8 & \checkmark & 15  & \checkmark  & 2.0 & \checkmark & 2.0 & \checkmark \\
\verb|kundu2_BUG|         & .02  & -          & 120 & -          & 1.6 & \checkmark & 1.9 & \checkmark & 510 & \checkmark &  1.6 & \checkmark & 1.6 & \checkmark \\
\verb|pc_sfifo_1|         & .02  & -          & 12  & -          & 14  & -          & 900 & -          & 5.3 & \checkmark  & 19  & \checkmark & 19  & \checkmark \\
\verb|pc_sfifo_2|         & .02  & -          & 6.3 & -          & 14  & -          & 900 & -          & 9.8 & \checkmark &  24  & \checkmark & 24  & \checkmark \\
\verb|token_ring.01|      & .02  & -          & 16  & -          & 1.7 & -          & 1.6 & -          & 8.1 & \checkmark  & 9.8 & \checkmark & 9.8 & \checkmark \\
\verb|toy2_BUG|           & .03  & -          & 30  & -          & 2.7 & \checkmark & 2.1 & \checkmark & 65  & \checkmark  & 2.7 & \checkmark & 2.7 & \checkmark \\
\verb|transmitter.01.BUG| & .02  & -          & 5.3 & \checkmark & 1.7 & \checkmark & 1.5 & \checkmark & 2.2 & \checkmark  & 1.7 & \checkmark & 1.7 & \checkmark \\
\verb|transmitter.02.BUG| & .02  & -          & 21  & \checkmark & 2.0 & \checkmark & 1.7 & \checkmark & 5.1 & \checkmark  & 2.0 & \checkmark & 2.0 & \checkmark \\
\verb|transmitter.03.BUG| & .03  & -          & 92  & -          & 2.0 & \checkmark & 1.8 & \checkmark & 36  & \checkmark  & 2.0 & \checkmark & 2.0 & \checkmark \\
\verb|transmitter.04.BUG| & .03  & -          & 270 & -          & 2.2 & \checkmark & 2.5 & \checkmark & 900 & -           & 2.2 & \checkmark & 2.2 & \checkmark \\
\verb|transmitter.05.BUG| & .04  & -          & 670 & \checkmark & 2.5 & \checkmark & 2.3 & \checkmark & 900 & -           & 2.5 & \checkmark & 2.5 & \checkmark \\
\verb|transmitter.06.BUG| & .04  & -          & 900 & -          & 3.4 & \checkmark & 3.0 & \checkmark & 900 & -           & 3.4 & \checkmark & 3.4 & \checkmark \\
\verb|transmitter.07.BUG| & .04  & -          & 900 & -          & 3.8 & \checkmark & 3.3 & \checkmark & 900 & -           & 3.8 & \checkmark & 3.8 & \checkmark \\
\verb|transmitter.08.BUG| & .05  & -          & 900 & -          & 6.3 & \checkmark & 4.6 & \checkmark & 900 & -           & 6.3 & \checkmark & 6.4 & \checkmark \\
\verb|transmitter.09.BUG| & .05  & -          & 900 & -          & 11  & \checkmark & 9.4 & \checkmark & 900 & -           & 11  & \checkmark & 11  & \checkmark \\
\hline \T \B \bfseries SystemC total & 0.49 & 0 & 5000 & 3 & 88 & 13 & 1900 & 13 & 7200 & 10 & 320 & 17 & 320 & 17\\\hline\hline  \T 
\bfseries Total & 4.4 & 9 & 5100 & 15 & 230 & 35 & 3300 & 37 & 7800 & 37  & 790 & 44 & 770 & 44

\end{tabular}
\caption[]{
\begin{minipage}[t]{11.6cm}
Comparison of different model-checking configurations and their combinations\\
\normalfont
Combination A: Explicit with 10\,s time limit and Predicate\\
Combination B: CBMC with $k=1$, Explicit with 10\,s time limit and Predicate\\
The run-time values are given in seconds and with two significant digits.
\end{minipage}}
\label{tab:combinations}
\vspace{-10mm}
\end{table*}

\smallsec{Configurations}
We experiment with three verification techniques and their combinations.
First, we run \cbmc with two different values for the loop bound~$k$
(columns `CBMC $k=1$' and `CBMC $k=10$').
Second, we use \cpachecker with explicit-value analysis
with a time bound of 10\,s (`Explicit $time(10s)$'), and without a time bound (`Explicit').
Third, we run \cpachecker with predicate analysis (`Predicate').
These first five data columns of Table~\ref{tab:combinations} 
report the results of using such configurations.

The last two columns show the results for two combinations of model-checking configurations.
Different verification tools and techniques perform better on different verification
problems based on their focus and capabilities.
For example, some configurations are better in detecting shallow bugs and some are
better in proving the absence of safety violations.
We combine them by running the first configuration, which is possibly limited by some condition.
If this analysis terminates with a result of `bug found' or `program is safe',
the verification is finished.
If the analysis, however, terminates without a final verification result,
e.g., due to imprecision or due to the condition preventing a complete analysis,
we will restart the analysis with the next configuration,
until a final result is obtained or there is no further configuration.
In all cases, the final result is the result of the last configuration,
and the total time is the sum of the run-times of all used configurations.

\smallsec{Discussion}
\cbmc with the loop bound set to 1 is not able to verify many programs,
but is a very fast configuration.
The total time necessary to analyze all 44 programs is only about 3.4\,s.
It should be noted that when \cbmc is not able to produce a useful result,
the run-time is almost always under 0.1\,s.
Although \cbmc is a bounded model checker,
it is able to verify the safety of a few programs that contain only simple or no loops.
When increasing the loop bound (e.g., to $k=10$), \cbmc is able to verify some more programs,
but also needs much more time.
This even happens for programs with bugs, where it reaches the timeout in several cases.
Much time and effort is spend to verify only six additional programs (cf. rows `total').

The explicit-value analysis of \cpachecker (limited with $time(10s)$) is able 
to find many more bugs than \cbmc with $k=10$,
while using significantly less time.
In contrast to \cbmc, this configuration spends more time on examples for which it is not able to
compute a verification result.
The reason is that it discovers one abstract state after the other,
terminating as soon as an error state has been reached
or the program was verified completely.
However, the run-time is still only a few seconds for all examples,
and it is much less than \cbmc with loop bound~10,
although the number of successfully verified programs is significantly higher.
For some programs (e.g., for \verb|diskper_simpl1|),
the model checker runs a few seconds longer than the time limit.
This is because our implementation of the time limit affects only the actual analysis,
but the time listed in the table is the total run-time,
including program startup and parsing.

The two columns `Explicit' (without condition) and `Predicate' show the results
of running \cpachecker with explicit-value analysis and predicate analysis, respectively.
These configurations are bounded only by the global time limit of 15\,min,
no condition is used.
Due to this fact they are able to verify a substantial number
of programs, compared to the bounded configurations.
However, both fail on several examples due to a time out,
and thus have a dramatically high total run-time.
Explicit analysis is sometimes too imprecise to verify a safe program
(e.g., {\small \verb|bist_cell|}).

A comparison of the columns `Explicit' and `Explicit $time(10s)$'
also demonstrates how conditions can be useful in optimizing the verification process.
As shown in the row `Total' at the bottom of the table, with time-limit condition it takes a total of 230\,s
to run all programs whereas it takes 3300\,s without the condition.
Even though the total run-time is significantly higher, explicit analysis without condition can successfully analyze only two more programs.
The configuration using a condition also reports the assumptions generated during the analysis so that the user can 
see which parts of the state space were analyzed and which were not.

By combining several configurations,
the effectiveness of the verification process can be improved significantly,
both in the number of verified programs as well as in the run-time.
Bounded model-checking is good in finding shallow bugs, 
and explicit analysis with the limiting condition is good in finding possibly deep 
but still ``easy'' bugs,
but for proving program safety, predicate abstraction is the best choice.
Thus, we first spend a limited effort in finding bugs with the efficient configurations.
If no final result is given, we switch to the next more powerful ---and more expensive--- analysis.

Explicit-value analysis can efficiently identify some bugs that predicate abstraction
cannot find, even if the former is used as a conditional model checker.
If the explicit-value analysis with a time-limit condition 
does produce a useful result, then it is usually faster than predicate abstraction,
although this configuration is not powerful enough to analyze some programs due
to its limitations.
The combination of these two configurations together is strictly more powerful
than any configuration alone, and can verify all benchmark programs, but is also more expensive
in terms of run-time.

\begin{table*}[t]
\centering
\footnotesize
\newcommand\smallerfontsize\tiny
\begin{tabular}{l|D{.}{.}{3.0}c|D{.}{.}{5.0}c|D{.}{.}{5.0}c|D{.}{.}{4.1}cc}
Program & \multicolumn{2}{c|}{Explicit} & \multicolumn{2}{c|}{Predicate} & \multicolumn{2}{c|}{Comb. A} & \multicolumn{2}{c}{Conditional MC} \\
& & & & & \multicolumn{2}{c|}{Explicit + Predicate} & \multicolumn{2}{c}{Explicit + Predicate} \\
\hline
\verb|token_ring.01.BUG| & 1.6 & - & 4.0 & \checkmark & 5.6 & \checkmark & 2.6 & \checkmark \\
\verb|token_ring.01|     & 1.9 & - & 8.9 & \checkmark & 11  & \checkmark & 2.4 & \checkmark \\
\verb|token_ring.02.BUG| & 2.3 & - & 24  & \checkmark & 26  & \checkmark & 4.3 & \checkmark \\
\verb|token_ring.02|     & 1.7 & - & 900 & -          & 900 & -          & 4.0 & \checkmark \\
\verb|token_ring.03.BUG| & 2.6 & - & 900 & -          & 900 & -          & 5.4 & \checkmark \\
\verb|token_ring.03|     & 2.5 & - & 900 & -          & 900 & -          & 5.1 & \checkmark \\
\verb|token_ring.04.BUG| & 2.7 & - & 900 & -          & 900 & -          & 9.1 & \checkmark \\
\verb|token_ring.04|     & 2.5 & - & 900 & -          & 900 & -          & 8.5 & \checkmark \\
\verb|token_ring.05.BUG| & 3.8 & - & 900 & -          & 900 & -          & 16  & \checkmark \\
\verb|token_ring.05|     & 3.2 & - & 900 & -          & 900 & -          & 17  & \checkmark \\
\verb|token_ring.06.BUG| & 4.8 & - & 900 & -          & 900 & -          & 34  & \checkmark \\
\verb|token_ring.06|     & 5.4 & - & 900 & -          & 900 & -          & 40  & \checkmark \\
\verb|token_ring.07.BUG| & 9.1 & - & 900 & -          & 900 & -          & 140 & \checkmark \\
\verb|token_ring.07|     & 8.3 & - & 900 & -          & 900 & -          & 180 & \checkmark \\
\verb|token_ring.08.BUG| & 25  & - & 900 & -          & 900 & -          & 580 & \checkmark \\
\verb|token_ring.08|     & 6.0 & - & 900 & -          & 900 & -          & 720 & \checkmark \\
\verb|token_ring.09.BUG| & 120 & - & 900 & -          & 900 & -          & 900 & -          \\
\verb|token_ring.09|     & 130 & - & 900 & -          & 900 & -          & 900 & -          \\
\verb|mem_slave_tlm.1|   & 2.0 & - & 900 & -          & 900 & -          & 5.2 & \checkmark \\
\verb|mem_slave_tlm.2|   & 2.8 & - & 900 & -          & 900 & -          & 6.3 & \checkmark \\
\verb|mem_slave_tlm.3|   & 3.0 & - & 900 & -          & 900 & -          & 7.5 & \checkmark \\
\verb|mem_slave_tlm.4|   & 3.4 & - & 900 & -          & 900 & -          & 8.1 & \checkmark \\
\verb|mem_slave_tlm.5|   & 3.8 & - & 900 & -          & 900 & -          & 10  & \checkmark \\
\verb|toy|               & 2.6 & - & 900 & -          & 900 & -          & 7.6 & \checkmark \\
\hline \T \B \bfseries Total & 350 & 0 & 19000 & 3 & 19000 & 3 & 3600 & 22 \\
\end{tabular}
\caption{Experiments with assumption automata as condition}
\label{tab:experiments-automaton}
\vspace{-10mm}
\end{table*}

\cbmc with $k=1$ is always extremely fast to compute a verification result, 
or it gives up after a negligible amount of time (ca. 0.1\,s) such that
the effort spent in \cbmc does not negatively influence the overall verification result.
Since \cbmc performs better on simplified NT device drivers, it improves the overall result.
Each of the three tools and configurations has unique advantages,
so it makes sense to combine all three tools and configurations together
via conditional model checking, and the results are reported in the last column (Combination~B).
This combination can solve all instances that Combination~A can solve,
but in addition, the performance is slightly increased for those programs that \cbmc can solve.
\cbmc with a very low loop bound produces practically no overhead,
and thus, this configuration should always be included.
In conclusion, via conditional model checking, we first run an inexpensive shallow analysis
to identify the easy bugs, then run a slightly more expensive, more in-depth analysis
to solve more instances, and last run the most expensive analysis.

\subsection{Model Checking using Conditions from Previous Runs}

Now we demonstrate how we can utilize the output condition (assumption)
of one conditional model checking to improve the results of a successive conditional model checking,
by feeding the negation of the generated condition of the first run as input condition for the second run.
(This is currently restricted to different configurations of the tool \cpachecker
because \cbmc does not (yet) support the output or/and input of conditions.)

\smallsec{Configurations}
We experiment with explicit analysis and predicate analysis.
The first column of the Table~\ref{tab:experiments-automaton} reports results for explicit analysis,
the second one the results for predicate analysis with LBE.
The third column combines explicit analysis and predicate analysis
as we did in Table~\ref{tab:combinations}.
The last column combines these two configurations in the following way:
First, explicit analysis with the Assumption Storage CPA is run
and an assumption automaton based on the generated assumptions is produced.
This automaton is similar to the automaton for the example shown in Fig.~\ref{fig:example-non-linear-Automaton}.
Second, predicate analysis is started using this automaton as the condition to restrict the searched state space.
The state space that did not remain unchecked by explicit analysis is pruned using this automaton.
This way we hope that predicate analysis will run the analysis on a smaller part of the state space
and more programs can be checked.

\smallsec{Discussion}
For these programs, explicit analysis finds only counterexamples which are verified as infeasible by CBMC
and thus it reports `unknown' after spending some time (up to 2\,min) and nothing can
be concluded about these programs.
Predicate analysis reports a correct result for three programs but it times out for the remaining 21 programs
and consumes a significant amount of time.
For a majority of these programs, combining explicit analysis and predicate analysis does not
produce a good result as well.
This is because both analysis fails to return an answer for the given program.
In other words, explicit analysis can run all the programs in only 350\,s and visits hundreds of thousands of abstract states,
but is too imprecise, whereas predicate analysis is precise enough but too expensive.
The last column shows a dramatic improvement
when both analyses are combined using conditional model checking,
where the assumption produced by explicit analysis is used
for restricting the state space analyzed by predicate analysis.
Almost all of these programs can be successfully analyzed.
Even though the combined configuration consumes significantly less time than predicate analysis alone,
19 more programs can be verified.
In most cases, programs can be checked in less than 1\,min.

\section{Conclusion}

Software model checking is an undecidable problem,
and therefore, we cannot create model checkers that always
give a precise answer to the verification problem.
Conventional model checkers fail when they cannot give
a precise answer, leaving the user no information about
what the tool was not able to verify,
where in the program the problem occurred, and
how much of the program was already verified.
Conditional model checking proposes to design model checkers
that do not fail, but instead summarize their work
when they decide to give up.
That is, we change the outcome from \{safe, unsafe, fail\} to \{condition\},
meaning that the model checker has verified that
the program satisfies the specification under the reported condition.

In addition to this user feedback,
conditional model checking improves \emph{verification coverage}
and \emph{performance}
by making it possible to give
conditions as input that avoid certain parts
of the program.
Our experiments show that such conditions significantly improve
the behavior of the model checker for programs that violate the specification.
This was expected, because techniques such as bounded model checking,
which are limiting the loop unwinding during state-space exploration,
are known to be helpful for finding bugs.
Conditional model checking is a generalization of this concept, because it can use an arbitrary 
condition to prevent traversing certain parts of the state space.

The main benefit of conditional model checking is yet another one.
Conditional model checking can be used to combine different verification tools,
or different runs of the same tool with different conditions.
After the first verification run, the resulting assumption is given as input to the second verification
run, which uses a different algorithm or condition, and therefore can be expected to fail on
different parts of the state space.
This process can be repeated until the desired coverage is achieved.

We experimentally confirmed the benefits of conditional model checking, 
by running different tools and configurations one after the other.
The best setting was to first run a shallow analysis
(\cbmc with the condition to perform only one loop unrolling),
then, using the assumption that resulted from the first analysis, run an in-depth analysis
(\cpachecker with explicit-value analysis and a time limit of~10\,s),
and finally, using the assumption produced by the second analysis, run the exhaustive and most expensive algorithm
(\cpachecker with predicate analysis and the condition to report the verification assumptions
after 15\,min of run-time).

In summary, we have demonstrated that with conditional model checking:
(1)~More problem instances can be solved.
(2)~Performance can be improved, in terms of reduced run-time and reduced memory consumption.
(3)~Better coverage of the state space can be achieved for problem instances for which
neither the presence nor the absence of an error can be proved.
(4)~A powerful and flexible combination of different verification technologies is enabled.

In our future work, we will investigate
heuristics to algorithmically find a set of appropriate input conditions and the most promising
sequence of verification approaches,
such that the verification flow that we used in our experiments can
be determined automatically by the verification tool.
We also plan to investigate further applications of conditional model checking.
For example, conditional model checking can support the verification 
of components and modules in isolation, and 
then be used to compose the global verification goal of the system from the partial results
(i.e., modeling assumptions and guarantees as `conditions').

\balance

\end{document}